# DEPENDENCE OF CHLOROPHYLL CONTENT IN LEAVES FROM LIGHT REGIME, ELECTROMAGNETIC FIELDS AND PLANT SPECIES

Alexander Kholmanskiy, Nataliya Zaytseva

A.I.Yevdokimov Moscow State University of Medicine and Dentistry, Russia

*E-mail:* allexhol@ya.ru, https://orcid.org/0000-0001-8738-0189)

**Abstract**

*The regularity of the distribution of chlorophylls content in a series of 30 cultivated plants and 75 steppe grasses was studied. The increased content of chlorophyll and magnesium in vegetables and grains compared with greens and steppe grasses is associated with more complex genetics of metabolism, which has stages of flowering and fruiting. The chlorophyll content increases with the use of LED phytoirradiators with an emission band coinciding with the first absorption band of chlorophyll. Industrial electromagnetic fields can affect the biosynthesis of pigments in deciduous trees, but cultivated herbaceous plants are not sensitive to them.*

**Introduction**

The productivity of photosynthesis of green plants is determined mainly by their genetics and strongly depends on temperature, nutrient medium and light regime (Polevoj, 1989 [25]; **Andrianova and Tarchevskij**, 2000 [2]). These factors determine the structure and efficiency of the photosynthetic apparatus of the plant (PSA). The key functional elements of PSA are **a** and **b** forms of the chlorophyll (Chl) (Senge et al., 2014 [28]; Kholmanskiy and Smirnov, 2018 [16]) and water, which plays the role of a dynamic matrix and active metabolite (Polevoj, 1989 [25]; Kholmanskiy and Tilov, 2013 [13]; Aksenov, 2004 [1]). Water sensitizes PSA by forming complexes with Mg and chiral centers Chl (Kholmanskiy and Smirnov, 2018 [16]) and other molecules in the composition of PSA (Senge, et al. 2014 [28]; Kholmanskiy, 2016 [14]; Kholmanskiy, 2018 [15]; Zaytseva and Sitanskaya, 2016 [37]). The abnormal thermodynamic properties of water optimize the physics of cold stratification of seeds at a temperature of about 4 °C (Kholmanskiy, 2018 [15]) and minimize the energy of seed germination and photosynthesis in the temperature range 15-25 °C (Polevoj, 1989[25]; Kholmanskiy, 2018 [15]). The photo- and thermophysics of Chl substantially depend on the Mg ion in the center of porphyrin cycle and the electronic nature of substituents in it (Senge, et al. 2014 [28]; Kholmanskiy and Smirnov, 2018 [16]). In principle, chlorophyll can play the role of a marker of the total Mg content in a plant and therefore can be considered an adequate characteristic of the nutritional value of a plant product.

The photochemistry and photophysics of PSA are determined primarily by the electronic structure of the ground and excited electronic states of Chl. A significant difference in the biophysics of chlorophyll Chl **a** and Chl **b** (Tyutereva and Ivanova, 2014 [35]) is caused by the replacement of the $CH_3$ group by the electron-acceptor and proton donor groups of CHO in the 7th position of pheofetin (Fig. 1). In (Frese et al., 2003 [9]; Kholmanskiy and Smirnov, 2018 [16]), the electronic nature of the ground and lower excited states of Chl was attributed to states with charge transfers whose dipoles are oriented along the mutually orthogonal *X* and *Y* axes of the Chl molecule. The high dipole moments of the excited states of Chls initiate electron and proton transfers from other PSA components and these reactions can accelerate the kinetics of photochemical reactions in PSA. It was suggested in (Kholmanskiy et al., 2019b [18]) that micropolarization of PSA promotes proton diffusion in leaves and intensifies the extraction of micro and macrocell ions by a plant.

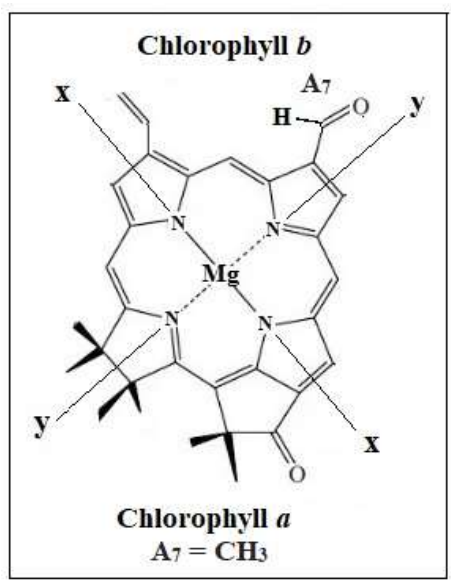

**Figure. 1.** Structures of the porphyrin cores of Chl*b* and Chl*a*.

Mg and Chl greatly increase the nutritional and medicinal value of cultivated plants, both leafy and fruit-bearing. In (Kholmanskiy et al.,, 2019a [17]; 2019b [18]) it was established that the efficiency of Mg extraction by plant leaves of cultivated plants depends on their species and responds to changes in the irradiation spectrum. It can be assumed that a similar dependence should be observed for the content of Chl in plant leaves.

To deepen understanding of the biophysical relationship between plant productivity and Chl we compared the total content of **a**+**b** Chl and Mg in raw samples of leaves of a number of cultivated plants and steppe grasses. We analyzed the dependence of the Chl content in the leaves of cultivated plants on the light regime, and also compared the effects on photosynthesis of Chl **a** and Chl **b** in plantain and wheat seedlings of an electromagnetic industrial (Novichkova and Podkovkin, 2003 [23]; Shashurin, 2014 [27]) and an electrostatic vortex field (Therapeutic reel Mishina [33]).

**Results and discussion**

The empirical data on the content of chlorophylls **a** and **b** in cultivated plants and grasses were taken from books and works published by us and other authors. Table 1 shows the total content of Chl **a**+**b** (hereinafter [Chl]) and the ratio **a**/**b** for raw samples of plant leaves. The determination of [Chl] in the works was carried out according to the methods described in (Lichtenthaler and

Buschmann, 2001 [21]). In (Bohn et al, 2014 [3]) [Chl] was determined using liquid chromatography, and the Mg content in Chl was calculated based on its mass fraction of ~2.7% in **a** and **b** Chl. In this work, we compared the concentrations of total Mg and Mg in the composition of Chl (Table 2).

**Table 1. The content of chlorophyll a and b in the leaves of different plants**

| Sample | | Time growth – week (w); irradiation conditions – Mercury Lamp (ML); Sodium Lamp (SL); LED; Sun | | Chlorophyll (mg/g) | | Ref. |
|---|---|---|---|---|---|---|
| | | | | a+b | a/b | |
| Tomato | | 5 varieties | | 1.2 | 2.7 | (Kavtsevich, 2015 [11]) |
| | | SL | | 1.4 | 1.3 | (Smirnov and Kholmanskiy, 2017 [30]) |
| | | LED1 | | 1.7 | 1.1 | |
| | | Sun | | 1.8 | 1.5 | |
| Eggplant | | Extract in acetone | | 1.7 | - | (Mezhunts, 2009 [22]) |
| | | | | 1.8 | | |
| Cucumber | | | | 1.8 | | |
| | | ML→ 2ML | | 1.6→0.9 | 3 | (Dalke, 2014 [7]) |
| | | SL | | 0.8 | 1.7 | (Smirnov, 2018 [31]) |
| | | LED1 | | 1.2 | 1.6 | |
| | | LED2 | | 1.5 | 1.4 | |
| Wheat | | 1 w | | 1.1 | 2.6 | (Liu, 2010 [20]) |
| | | 2 w | | 0.9 | 3.7 | |
| | | 3 w | | 1.5 | 3.7 | |
| Oats | | Sun | | 0.9 | - | (Petukhov, 2017 [24]) |
| Barley | | | | 3.5 | | Kushnareva and Perekrestova, 2015 [19]) |
| Peas | | | | 2.4 | | |
| Strawberry | | LED | | 2.1 | - | (Yakovtseva, 2015 [36]) |
| | | ML | | 1.6 | | |
| **Mean** | | | | **1.6±0.8** | **2.2±0.8** | |
| Lettuce | Afitsion | ML | 3 w | 0.4 | 4.6 | (Dalke, 2013 [13]) |
| | | | 5 w | 0.35 | 3.8 | |
| | Antony | SL, LED1, LED2 | 5 w | 0.7 | 2 | (Smirnov, 2018 [31]) |
| | Grand Rapids | LED ( Rt+FRt) | 2÷3 w | 1.3 | - | (Brazaitytė, 2006 [4]) |
| | | | 4 w | 0.5 | | |
| | | LED (Bl+Rt+FRt) | 2÷4 w | 0.8 | | |
| | Prickly lettuce | - | | 0.39 | 3.4 | (Bohn, 2004 [3]) |
| | Lettuce | | | 0,24 | 4.3 | |
| | Spinach | | | 0.79 | 2.5 | |
| | Rocket salad | | | 0.41 | 3.6 | |
| | Parsley | | | 0.63 | 3.1 | |
| | Cress | | | 0.31 | 2.6 | |
| | | ML, SL | | 1.4 | - | (Dalke, 2013 [6]) |
| Dill, parsley, cilantro, celery | | | | 1.8÷ 2 | | |
| Basil, arugula | | | | 1÷1.3 | | |
| **Mean** | | | | **1±0.5** | **3.3±0.7** | |
| Steppe plants, 67 species | | | | 0.85 | 1.8÷2.8 | (Ivanov, 2013 [10]) |
| Coltsfoot, red clover, mouse peas | | Sun | | 0.9 | 1.5÷1.6 | (Petukhov, 2017) |

| | | | 0.8 | | |
|---|---|---|---|---|---|
| Chamomile, meadow bluegrass | | | | | |
| Tutsan | | | 1.4 | 1.8÷2.5 | (Report, 2011 [26]) |
| Motherwort | | | 1.9 | | |
| Echinacea | | | 1.1 | | |
| **Mean** | | | **0.9±0.5** | **2±0.5** | |

**Table 2. The magnesium content in the raw leaf, fruit and chlorophyll**

| Sample (raw) | | Mg [8] | [Chl] | [Mg-Chl] | Mg/[Mg-Chl] |
|---|---|---|---|---|---|
| | | mg/g | | µg/g | |
| Lettuce | Leave | 0.14 | 0.7 | 20 | 7 |
| Tomato | | 1.1 | 1.5 | 40 | 28 |
| Cucumber | | 1.1 | 1.2 | 32 | 34 |
| | Fruit | 0.2 | 0.04 (Bohn, 2014 [3]) | 1 | 200 |

A large scatter of [Chl] values in different works is caused by differences in both their measurement methods and the growing conditions of the same plants. The [Chl] values determined for dry samples were recalculated for the wet state using a drying coefficient (k). It was evaluated by determining the weight of a fresh sample and after drying it at a temperature of no higher than 40 °C. For vegetable crops (tomato, cucumber), leaf (lettuce, greens) and steppe grasses their k were: 8.5; 15-20 and 6-7, respectively. At the same time, the proportion of water in raw samples was determined by the ratio (k – 1)/k and was equal to: 88%; 93-96% and 85-86%, respectively.

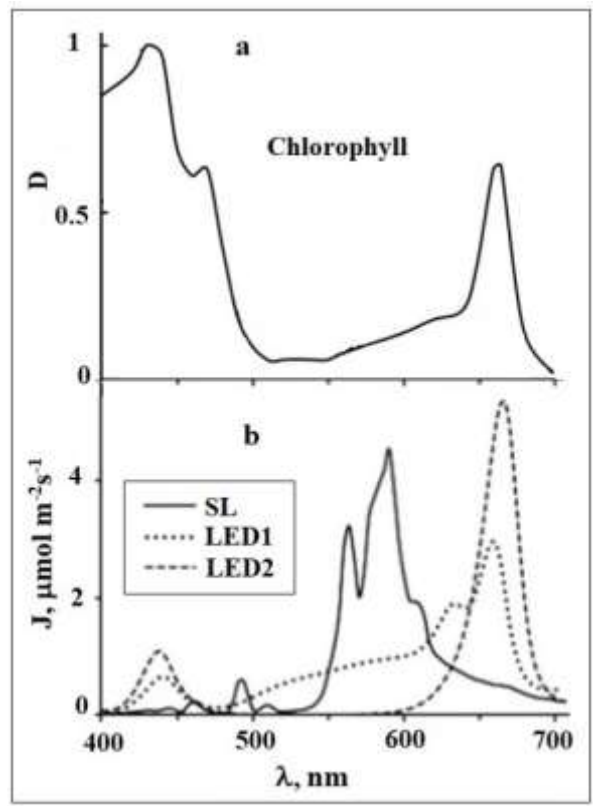

**Figure 2.** The absorption spectrum of a solution of Chl in ethanol (a) and the emission spectra of a sodium lamp (SL) and irradiators from two LED compositions, LED1 and LED2 (b). The initial spectra from (Kholmanskiy and Smirnov, 2018 [16]).

The Table 1 shows [Chl] values for plants growing in natural and greenhouse conditions. In the latter case, the age of the samples (weeks) and the light regimes were varied. The irradiators included various combinations of blue (Bl), green (Gr), red (Rt) light emitting diodes (LED1 and LED2) (Kholmanskiy et al, 2019b [18]) and 400 W high pressure sodium (SL) and mercury (ML) lamp. Their emission spectra and the absorption spectrum of Chl are shown in Figure. In some works, the LED composition included a diode emitting in the far-red region of the spectrum (FRt) with a maximum at 730 nm.

Plant species in Table 1 are divided into three groups: vegetables, cereals, berries (I); lettuce and greens (II); grasses (III). The mean values of [Chl] in the II and III groups are close and 1.6 times less than in the I group. At the same time, [Chl-**b**] in the I-st III-th is 2 times less than [Chl-**a**], and in the II-th one 3.3 times. The value of [Chl] in the first and second groups of plants reaches a maximum at 3 weeks of growth and depends on the spectral composition and intensity of irradiation. The Chl content decreases when the irradiation intensity is exceeded by the norms of the optimal light regime (Dalke et al, 2014 [7]). Moreover, the Chl photodestruction reaction can contribute to the negative effect (Kholmanskiy and Smirnov, 2018 [16]). The efficiencies of Chl biosynthesis in the leaves of plants of group I are close when the plants are irradiated with LED irradiators and the sun and are higher than when irradiated with SL. This result can be explained by the good overlap of the Rt emission bands of LED1 and LED2 with the first absorption band of Chl a and b (Fig. 2). In group II, the dependence of the efficiency of Chl biosynthesis on the irradiation spectrum is less pronounced.

Chl biosynthesis is limited by the efficiency of the plant's extraction of the Mg, which depends on the type of plant and the irradiation spectrum (Kholmanskiy et al., 2019b [18]). Table 2 shows the total content of Mg, Chl, and Mg in Chl ([Mg-Chl]) in the leaves and fruits of plants of the first and second groups, for samples grown under irradiation ML. From these values, the concentration ratio of total Mg to Mg in the composition of Chl (Mg/[Mg-Chl]) was calculated.

The obtained relations Mg/[Mg-Chl] indicate that the total Mg content in plants significantly exceeds [Chl], and this disproportion in plants of the 1st group is 4-5 times greater than in plants of the 2nd group. This is due to the fact that Mg is included in the active center of the enzyme, providing assimilation of $CO_2$ (Polevoj, 1989 [25]) and also participates in the activation of many other complexes (Senge et al, 2014 [28]; Shkol'nik, 1974 [29]; Sukovataya et al, 2008 [32]). In addition, the genetics of cucumber and tomato, unlike lettuce, includes programs for the stages of flowering and fruiting with their specific biochemistry and bioenergy (Shkol'nik, 1974 [29]; Tikhomirov, Sharupich, 2000 [34]).

A significant contribution to the intensification of photosynthesis and enzymatic reactions in animal and plant organisms is made by the magnetic isotope $^{25}$Mg (Buchachenko, 2014 [5]) the content of which in natural magnesium is 10%. Apparently, $^{25}$Mg in the composition of enzymes plays an important role in the functioning of the phytochrome and cryptochromic PSA systems, activating magnetically sensitive dark reactions of radical ion pairs (Evans et al., 2013 [8]). Chl charge transfer states in dark reactions relax with the formation of long-lived excitons (Frese et al., 2003 [9]). Their participation in the work of PSA can, in principle, determine the dependence of the efficiency of photosynthesis on external electromagnetic and electric fields. To verify this assumption, we compared the dependence of the content of pigments – Chl **a**, Chl **b** and carotenoids (**k**) in maple leaves growing at a distance of 0 to 1000 m (control) from 110 kV power lines (Novichkova and Podkovkin, 2003 [23]) and in the leaves of plantain seedlings, grown in an electromagnetic field (EMF) with a frequency of 50 Gz and different intensities (Shashurin et al., 2014 [27]), as well as wheat seedlings growing in an electrostatic vortex field with a frequency of 300 kGz (Therapeutic reel Mishina [33]). Recent experiments were carried out by us, using similar methods (Kholmanskiy and Smirnov, 2018 [16]; Kholmanskiy et al, 2019b [18]). The results and experimental conditions are shown in Table 3. The errors in measuring the pigment content in all samples were about 10%.

**Table 3. Effect of electromagnetic fields on pigment photosynthesis**

| Sample (raw) | Conditions of options. Control conditions. | Chlorophyll and carotenoids (mg/g) | | | | | |
| --- | --- | --- | --- | --- | --- | --- | --- |
| | | Options | | | Control | | |
| | | a+b | a/b | k | a+b | a/b | k |
| Wheat | Whirlwind electric field. Contr. – without field | 1.32 | 3 | 0.22 | 1.31 | 3 | 0.23 |
| Plantain | EM field 50 Gz (E=230-1800 V/m; B=350-2000 nTs). Contr. (E=12 V/m; B=50 nTs) | 0.81 | 1.7 | 0.66 | 0.84 | 1.6 | 0.62 |
| Maple | EM field power lines at 0-230 m. Contr. – at 1000 m | 0.62 | 2.6 | 0.14 | 0.45 | 2.5 | 0.1 |

From the data of Table 3 it follows that the influence of EMF affects the efficiency of pigment biosynthesis only in a tree. The reason for this may be the elongation of transport communications along which charged mineral elements move from bottom to top, and phytohormones from top to bottom. The electrophysical properties of the sap in the layers of cambium and sapwood of deciduous trees, in contrast to the resin of conifers, contribute to electrotropism (Kholmanskiy, 2009 [12]).

**Conclusion**

Thus, it was found that the content of chlorophyll and magnesium in the leaves of cultivated plants is significantly higher in fruitful species than in leafy and, especially in steppe grasses. The chlorophyll content increases when growing plants in light regime using LED irradiators having a radiation band that overlaps well with the first absorption band of chlorophyll **a** and **b**. Differences in the content of chlorophyll and plant productivity are associated with a more complex genetic program of metabolism in higher cultivated plants, including the stages of flowering and fruiting.